# Phosphorene nanoribbon as a promising candidate for thermoelectric applications

J. Zhang[1], H. J. Liu[1,*], L. Cheng[1], J. Wei[1], J. H. Liang[1], D. D. Fan[1], J. Shi[1], X. F. Tang[2], Q. J. Zhang[2]

[1]*Key Laboratory of Artificial Micro- and Nano-Structures of Ministry of Education and School of Physics and Technology, Wuhan University, Wuhan 430072, China*

[2]*State Key Laboratory of Advanced Technology for Materials Synthesis and Processing, Wuhan University of Technology, Wuhan 430070, China*

In this work, the electronic properties of phosphorene nanoribbons with different width and edge configurations are studied by using density functional theory. It is found that the armchair phosphorene nanoribbons are semiconducting while the zigzag nanoribbons are metallic. The band gaps of armchair nanoribbons decrease monotonically with increasing ribbon width. By passivating the edge phosphorus atoms with hydrogen, the zigzag series also become semiconducting, while the armchair series exhibit a larger band gap than their pristine counterpart. The electronic transport properties of these phosphorene nanoribbons are then investigated using Boltzmann theory and relaxation time approximation. We find that all the semiconducting nanoribbons exhibit very large values of Seebeck coefficient and can be further enhanced by hydrogen passivation at the edge. Taking armchair nanoribbon with width $N$=7 as an example, we calculate the lattice thermal conductivity with the help of phonon Boltzmann transport equation. Due to significantly enhanced Seebeck coefficient and decreased thermal conductivity, the phosphorene nanoribbon exhibits a very high figure of merit ($ZT$ value) of 4.4 at room temperature, which suggests its appealing thermoelectric applications.

Recently, the black phosphorous has become the focus of science community due to the synthesis [1, 2, 3] of its two-dimensional form, namely, the phosphorene. Preliminary but exciting results indicate that phosphorene has potential application in nanoelectronics and optoelectronics. For example, the mobility of few layer phosphorene field-effect transistors (FET) can reach 286 cm$^2$/V/s and an on/off ratio of up to 10$^4$ at room temperature [1], which is comparable to that measured with phosphorene FET on Si/SiO$_2$ [2]. Li *el at.* found that the mobility of phosphorene based FET is thickness-dependent and can reach as high as 1000 cm$^2$/V/s with the thickness up to 10 nm [3]. On the other hand, theoretical calculations have been performed to predict the electronic and optical properties of the two-dimensional and one-dimensional phosphorene based materials. Dai *et al.* investigated the effect of stacking orders on the electronic properties of bilayer phosphorene and found the band gap of bilayer phosphorene can be changed from 0.78 to 1.04 eV by different stacking orders. Besides, they also suggested that the mixed phosphorene has potential application in thin-film solar cells [4]. By using first-principles simulations, Fei *et al.* showed that strain can control the anisotropic free-carrier mobility of monolayer phosphorene [5]. At the same time, Qiao *et al.* reported a systematic investigation on the electronic and optical properties of few-layer phosphorene. They found that the phosphorene is a direct band gap semiconductor, and the gap can be tuned from 1.51 eV of a monolayer to 0.59 eV of five layers. The mobility of hole and electron is anisotropic and can be as high as 7737 cm$^2$/V/s [6]. Guo *et al.* performed a comprehensive calculation to study the tensile stain and electric-field effect on the electronic properties of phosphorene nanoribbons, nanotubes, and van de Waals multilayers [7]. On the other hand, Tran *et al.* predicted that the band gap of differently oriented phosphorene nanoribbons show distinct scale laws with the variation of width [8]. Peng *et al.* discussed the electronic properties of phosphorene nanoribbons with different functional edges [9]. Apart from the rich electronic properties of low-dimensional phosphorous that may be applied in nanoelectronics and optoelectronics, a question is raised: can such a low-dimensional material be used for thermoelectric applications?

---

[*] Author to whom correspondence should be addressed. Electronic mail: phlhj@whu.edu.cn.

  In this work, we show by first-principles calculations that phosphorene nanoribbons (PNRs) with zigzag (ZPNRs) and armchair (APNRs) edges have distinct electronic properties, which can be further tuned by hydrogen passivation. Using the Boltzmann theory for both electrons and phonons, the transport coefficients of the PNRs are calculated. As a quick understanding of the thermoelectric performance of such kind of low-dimensional system, the *ZT* value of an APNR with width *N*=7 is evaluated and can be optimized to as high as 4.4 at room temperature. It is thus reasonable to expect that phosphorene nanoribbon could be a very promising candidate for high-performance thermoelectric applications.

The electronic structure properties of PNRs are performed by using the first-principles plane-wave pseudopotential formulation [10, 11, 12] as implemented in the Vienna *ab-initio* Simulation Package (VASP) [13]. The exchange-correlation energy is in the form of Perdew-Burke-Ernzerhof (PBE) [14] with generalized gradient approximation (GGA). The cutoff energy for the wave function is set to 400 eV. For the geometry optimization, a 1×1×15 Monkhorse-Pack *k*-meshes [15] is adopted for the Brillouin zone integration. The atomic positions are fully relaxed until the force on all atoms become less than 0.05 eV/Å. A vacuum distance of 14 Å is used for the directions of both width and thicknesses, so that the nanoribbons can be treated as independent entities. Based on the calculated energy band structure, the electronic transport properties are determined by using the semiclassical Boltzmann theory within the relaxation time approximation [16]. Such a method has been successfully used to predict some known thermoelectric materials, and the theoretical calculations agree well with the experimental results [17, 18, 19, 20]. In order to obtain reliable transport coefficients, a denser Monkhorse-Pack *k*-mesh up to 125 points in the irreducible Brillouin zone (IBZ) is used. To estimate the lattice thermal conductivity of PNRs, we use the phonon Boltzmann transport equation as implemented in the so-called ShengBTE code [21, 22, 23]. For the calculations of the second-order and third-order interatomic force constant matrix, we use a 1×1×6 and 1×1×3 supercell, respectively. The interactions up to the third nearest neighbors are considered when dealing with the anharmonic one. The *q*-point grid and scale parameter for Gaussian smearing are set as 1×1×35 and 1.0, respectively.

The PNRs can be obtained by cutting a monolayer phosphorene along armchair or zigzag directions. Following the conventional notation for graphene nanoribbon [24], the armchair phosphorene nanoribbons (APNRs) or zigzag phosphorene nanoribbons (ZPNRs) can be identified by the number of dimer lines or the zigzag chains across the ribbon width and are labeled as *N*-APNRs or *N*-ZPNRs, respectively. Fig. 1(a) and 1(c) show the ball-and-stick model of *N*-APNRs and *N*-ZPNRs, respectively. Here we consider *N*=7 ~ 12 for both armchair and zigzag nanoribbons, and the width varies from about 10 Å to 26 Å. Upon structure relaxations, we see from Fig. 1(b) and 1(d) there are some edge reconstructions for both APNRs and ZPNRs. Take *N*=9 as an example, For the APNR, we find that the bond length between the edge atoms on the translation direction decreases from 2.26 Å to 2.06 Å. The corresponding bond angles at edge increase from 96.9° to 111.0° for $\alpha$, and from 103.7° to 119.5° for $\beta$. In the case of ZPNR, the bond length at the edge decreases from 2.23 Å to 2.14 Å, and the corresponding edge angles $\gamma$ and $\theta$ increase from 96.9° to 100.7°, and from 103.7° to 108.9°, respectively. Similar edge reconstructions have been also found in previous calculations [9]. It should be mentioned that the reconstructed motifs at the opposite edges are symmetrically arranged, which is different from those found in the BiSb nanoribbon [25] with similar atomic configurations.

Fig. 2 plots the total energy of PNRs as a function of the ribbon width. We see that the ZPNRs are energetically more favorable than the APNRs. The total energy of both systems decrease monotonically with increasing ribbon width and has a tendency to approach that of monolayer phosphorene. It is known that monolayer phosphorene is semiconducting with a direct band gap of 1.51 eV [6]. By cutting the monolayer into nanoribbons, however, we find that the ZPNRs become metallic while APNRs remain semiconducting but exhibit an indirect band gap. As indicated in Fig. 3, the band gaps of APNRs decrease with increasing ribbon width, which can be attributed to the well-known quantum confinement effect. If the edge phosphorus atoms are passivated by hydrogen atoms, our calculated results indicate that the above-mentioned edge reconstructions disappear, and the band gaps of H-saturated APNRs become direct. Moreover, there are obvious increases of the corresponding band gaps (Fig. 3). Such effect is more pronounced for the ZPNRs, where a transition from metal to indirect band gap semiconductor is observed, and the H-passivated ZPNRs exhibit the largest band gaps among the four kinds of phosphorene nanoribbons. Our results are consistent with previous works [8, 9] and further confirm the reliability of our calculations.

We next consider the electronic transport properties of these nanoribbons by using the semiclassical Boltzmann theory and rigid-band approach [26]. Within this method, the chemical potential $\mu$ indicates the doping level (or carrier concentration) of the system, and the positive and negative $\mu$ correspond to *n*-type and *p*-type, respectively. The optimal carrier concentration can be obtained by integrating the density of states (DOS) of the system from the desired chemical potential to the Fermi level ($\mu$=0). Fig. 4(a) and 4(b) show the calculated Seebeck coefficients $S$ for the APNRs and H-passivated APNRs as a function of chemical potential at room temperature, respectively. We see that by optimizing the carrier concentration, the phosphorene nanoribbons can exhibit very large value of Seebeck coefficients. For example, the pristine 7-APNR has a maximum Seebeck coefficient of 1.36 mV/K at $\mu$=0.056 eV. Upon edge passivation, the maximum Seebeck coefficient can further enhanced 1.75 mV/K, which is significantly higher than most of previous results. As all the investigated ZPNRs in their pristine form are metallic, we see from Fig. 4(c) that their Seebeck coefficients are very small around the Fermi level. However, we observe a much higher Seebeck coefficient for the H-passivated ZPNRs (Fig. 4(d)), which can be optimized to 2.6 mV/K for the ZPNRs with width of $N$=7. Such giant Seebeck coefficients of phosphorene nanoribbons are very beneficial for their thermoelectric applications, and we will come back to this point later. On the other hand, we find from Fig. 4 that the Seebeck coefficients of three kinds of semiconducting nanoribbons exhibit obvious edge dependence, which is the largest for the H-passivated ZPNRs, the smallest for the pristine APNRs, with H-passivated APNRs in between. Such an order of Seebeck coefficients is a consequence of their distinct energy gaps. As shown in Fig. 5, the Seebeck coefficient at the peak shows linear dependence with the band gap and has no trend towards to saturation. This implies that the structures which have larger band gaps would have higher Seebeck coefficients. Similar relation has also been found in the graphene/h-BN superlattice nanoribbons [27].

Within the semiclassical Boltzmann theory, the electrical conductivity $\sigma$ can only be calculated with the relaxation time $\tau$ inserted as a parameter. This means that what we actually obtained is $\sigma/\tau$. To figure out the electronic relaxation time of phosphorene nanoribbons, we apply the deformation potential (DP) theory proposed by Bardeen and Shockley [28], where the relaxation time $\tau$ of 1D system at temperature $T$ can be expressed as:

$$\tau = \frac{\mu_c m^*}{e} = \frac{\hbar^2 C}{(2\pi k_B T)^{1/2} |m^*|^{1/2} E_1^2}. \tag{1}$$

In this formula, $m^*$ and $\mu_c$ are the effective mass and carrier mobility along the transport direction, respectively. The deformation potential constant $E_1 = \Delta E/(\Delta l/l)$ is determined by changing the lattice constant $l$ along the transport direction. The elastic modular can be obtained by $C = [\partial^2 E/\partial \delta^2]/l$, where $E$ is the total energy of the system, $\delta$ is the applied uniaxial strain along the ribbon direction.

For simplicity, in the following we take the APNRs with width of $N=7$ as an example to estimate the thermoelectric performance of phosphorene nanoribbons. Table I lists the calculated room temperature carrier mobility $\mu_c$, the relaxation time ($\tau$), and all the related parameters indicated in Eq. (1). Note that the calculated mobility for the electron is 168 cm$^2$/V/s at room temperature, which is smaller than the experiment values of two-dimensional phosphorene FET [1, 2], but close to that of monolayer MoS$_2$ [29]. The relaxation time for electron is higher than hole and comparable to that of bulk phosphorous along the armchair direction [30].

Fig. 6 (a) shows the calculated electrical conductivity $\sigma$ of 7-APNRs as a function of chemical potential $\mu$ at 300 K. We see $\sigma$ increases gradually with the increasing absolute $\mu$. However, the electrical conductivity is very small at the chemical potential ($\mu$=0.056 eV) where the Seebeck coefficient approach its peak value (See Fig. 4(a)). This implies that there should be a comprise between the Seebeck coefficient and the electrical conductivity. Indeed, we see from the Fig. 6(b) that the maximum power factor appears at $\mu$=0.43 eV, where neither the Seebeck coefficient nor the electrical conductivity reach the maximum. The calculated electronic thermal conductivity ($\kappa_e$) shown in Fig. 6(c) has the same behavior as that of electrical conductivity since $\kappa_e$ is calculated by the Wiedemann-Franz law [31]

$$\kappa_e = L\sigma T, \qquad (2)$$

where the Lorentz number is set to $2.45\times 10^{-8}$ W$\Omega$K$^{-2}$ [32, 33, 34].

To estimate the lattice thermal conductivity ($\kappa_p$) of phosphorene nanoribbons, the Boltzmann transport equation for the phonons are adopted [21, 22, 23]. The calculated lattice thermal conductivity for the 7-APNR is 0.102 W/mK at room temperature. Note the definition of a cross-sectional area has some arbitrariness for low-dimensional systems such as our phosphorene nanoribbons, and this value of thermal conductivity is calculated with respect to a vacuum distance of 14 Å at the directions of both width and thickness. If instead we use the interlayer distance of the bulk phosphorus (3.07 Å) and the real ribbon width of 10.44 Å, the lattice thermal conductivity is re-calculated to be 0.75 W/mK, which is much smaller than that of bulk value (12.1 W/mK) [35] and suggests that phosphorene nanoribbons may have better thermoelectric performance than their bulk counterpart. The reduced lattice thermal conductivity can be attributed to additional phonon scattering on the boundary (edge) of one-dimensional nanoribbons. Compared with the electronic lattice thermal conductivity (see Fig. 6(c) and the inset), the lattice part is relatively higher in the chemical potential range from −0.36 to 0.34 eV. However, the electronic part becomes dominative and increases rapidly beyond this range.

With all the transport coefficients available to us, we can now evaluate the figure of merit (*ZT* value) of phosphorene nanoribbons, which is given by

$$ZT = \frac{S^2 \sigma T}{\kappa_p + \kappa_e}. \tag{3}$$

Fig. 6(d) shows the calculated *ZT* value of 7-APNR as a function of chemical potential at room temperature. At appropriate carrier concentration (*p*-type with $\mu =$ −0.36 eV and *n*-type with $\mu =$ 0.35 eV) where the electronic and lattice thermal conductivity are similar to each other (see the inset of Fig. 6(c)), we find that the *ZT* value can be optimized to 2.2 and 4.4, respectively. Such *ZT* values significantly exceeds the performance of most laboratory results reported and are comparable to the efficiency of traditional energy conversion method, which makes phosphorene nanoribbon a very promising candidate for thermoelectric applications. As summarized in Table II, the effectiveness of the phosphorene nanoribbon is mainly due to its very large Seebeck coefficient and much lower thermal conductivity (especially the lattice part) compared with the bulk phosphorous [30]. It should be mentioned that although we only consider the thermoelectric performance of a particular phosphorene nanoribbons (7-APNRs), it is reasonable to expect that other kinds of phosphorene nanoribbons with different width and/or edge configurations can have similar thermoelectric performance, especially for those with edge passivation. Our theoretical work may serve as a guide for further experimental efforts using the phosphorene nanoribbons as a high-performing thermoelectric material.

In summary, we have demonstrated that phosphorene nanoribbons could be optimized to exhibit very good thermoelectric performance with a maximum *ZT* value of 4.4 at room temperature. Our theoretical calculations are self-consistent within the framework of density functional theory and Boltzmann theory, and there are no adjustable parameters. We want to mention that the phosphorous thin films and few-layer or even monolayer phosphorene were recently fabricated by using mechanical exfoliation method [1, 2, 3, 36]. It is reasonable to expect that by cutting monolayer phosphorene or by patterning epitaxially grown phosphorene, one can obtain the phosphorene nanoribbons with controllable width and/or edge, and thus realize their possible thermoelectric applications.

We thank financial support from the National Natural Science Foundation (Grant No. 51172167 and J1210061) and the "973 Program" of China (Grant No. 2013CB632502).

**Table I** The deformation potential $E_1$, stretching modulus $C$, effective mass $m^*$, mobility $\mu_c$, and relaxation time $\tau$ for hole and electron along periodic direction in the 7-APNR at 300 K.

| Carriers | $E_1$ (eV) | $C$ (eV/Å) | $m^*$ ($m_e$) | $\mu_c$ (cm$^2$V$^{-1}$s$^{-1}$) | $\tau$ (fs) |
|---|---|---|---|---|---|
| hole | −1.420 | 7.667 | 0.44 | 102.2 | 25.8 |
| electron | −0.826 | 7.667 | 0.65 | 168.8 | 62.9 |

**Table II** Maximum $ZT$ values and corresponding electron transport coefficients at optimized chemical potential (or carrier concentration $n$) for the 7-APNR at room temperature. Note for the one-dimensional PNR, the carrier concentration is calculated with respect to a vacuum distance of 14 Å at the directions of both width and thickness.

| Carrier type | $\mu$ (eV) | $n$ (e/uc) | $S$ (μV/K) | $\sigma$ (10$^4$S/m) | $S^2\sigma$ (W/mK$^2$) | $\kappa_e$ (W/mK) | $ZT$ |
|---|---|---|---|---|---|---|---|
| $p$-type | −0.36 | 0.01465 | 334.5 | 1.33 | 1.49×10$^{-3}$ | 0.10 | 2.2 |
| $n$-type | 0.35 | 0.00992 | −426.8 | 1.98 | 3.61×10$^{-3}$ | 0.15 | 4.4 |

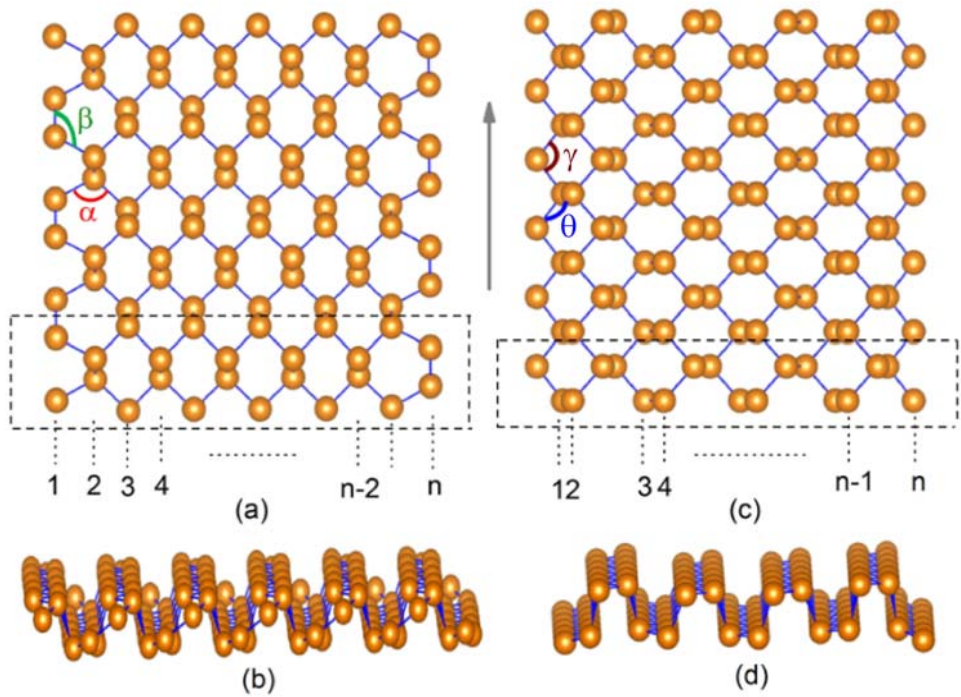

**Figure 1** The ball-and-stick model of armchair phosphorene nanoribbon: (a) is top-view, and (b) is side-view. The zigzag counterpart is plotted in (c) and (d), respectively. The unit cell is outlined by dash lines and the arrow indicates the translation direction of nanoribbons.

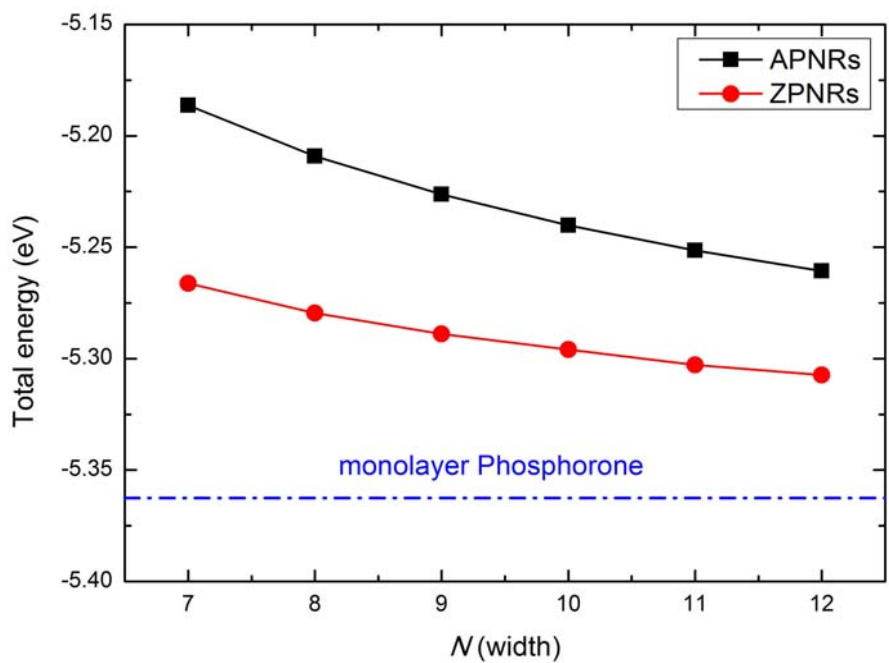

**Figure 2** The calculated total energies (in units of eV per atom pair) of the APNRs and the ZPNRs as a function of the ribbon width. The dotted line indicates the energy of the monolayer phosphorene.

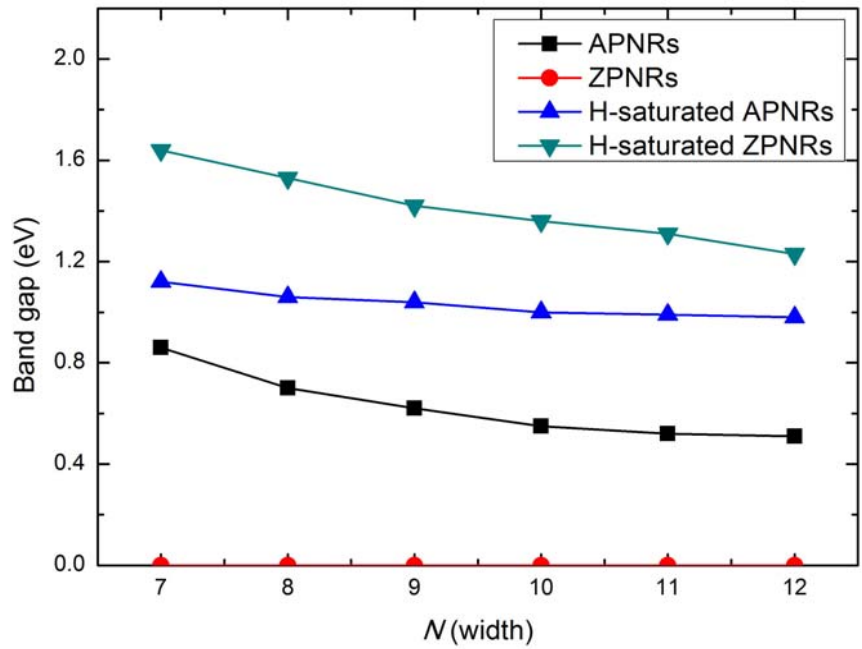

**Figure 3** The calculated energy band gaps for phosphorene nanoribbons as a function of width.

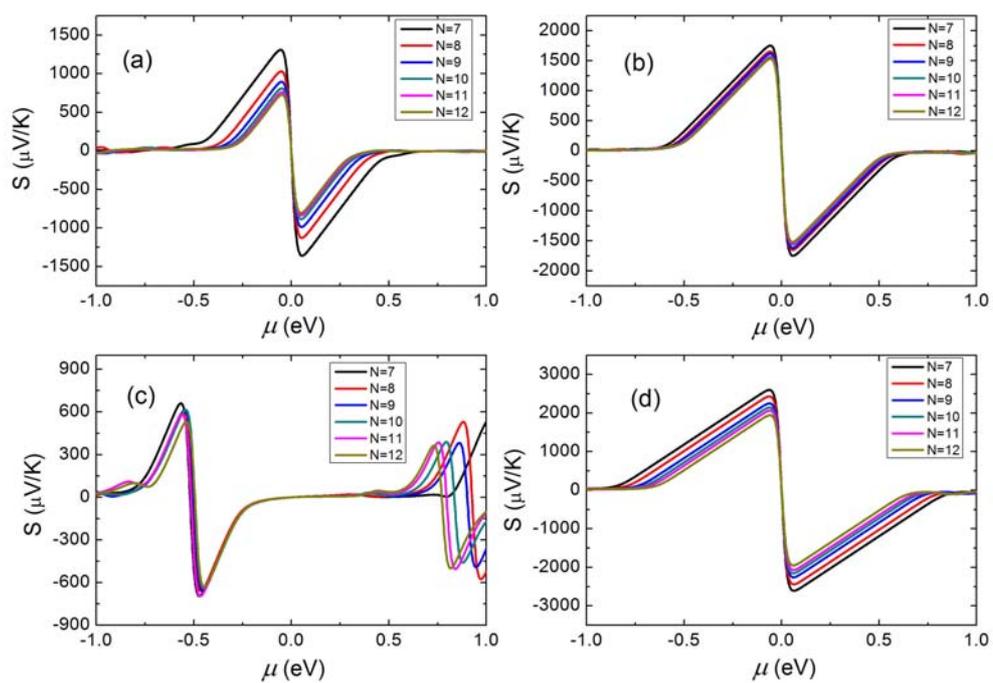

**Figure 4** The calculated Seebeck coefficient of (a) APNRs, (b) H-saturated APNRs, (c) ZPNRs, and (d) H-saturated ZPNRs as a function of chemical potential at 300K.

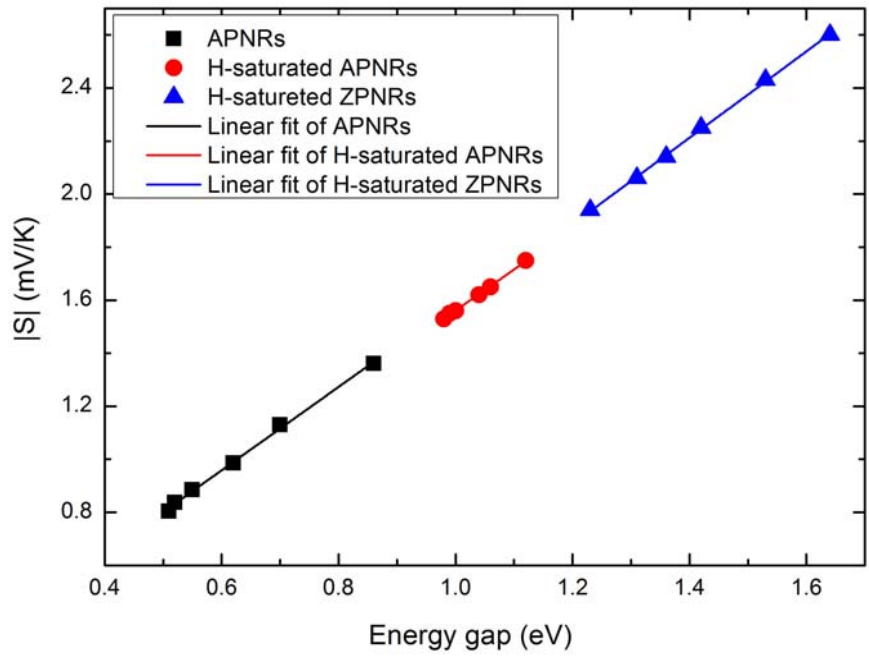

**Figure 5** The room temperature Seebeck coefficients (peak value) as a function of band gaps for APNRs, H-saturated APNRs, and H-saturated ZPNRs.

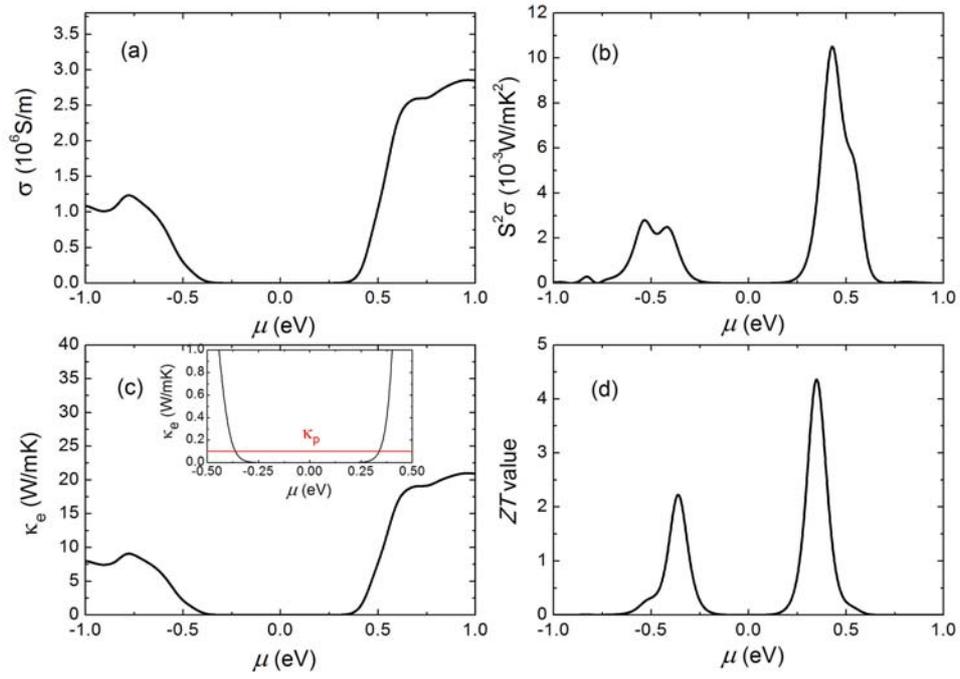

**Figure 6** The calculated electronic transport coefficients at 300 K as a function of chemical potential for the 7-APNRs: (a) the electrical conductivity, (b) the power factor, (c) the electronic thermal conductivity, and (d) the *ZT* value. The red line in the inset of (c) indicates the lattice thermal conductivity of 7-APNRs.